\documentclass[aps,prb,twocolumn]{revtex4}
\usepackage{graphicx}
\usepackage{amssymb}%
\usepackage{epsfig}
\usepackage{amsmath}
\usepackage{ulem}
\usepackage{color}
\usepackage{subfigure}

\begin{document}
\title{Heterogeneous freezing in a geometrically frustrated spin model without disorder: spontaneous generation of two time-scales}

\author{O.~C\'epas and B.~Canals}
\affiliation{Institut N\'eel, CNRS et Universit\'e Joseph Fourier, BP 166, F-38042 Grenoble 9, France }

\begin{abstract} 
By considering the constrained motion of classical spins in a
geometrically frustrated magnet, we find a dynamical freezing
temperature below which the system gets trapped in metastable states
with a ``frozen'' moment and dynamical heterogeneities. The residual
collective degrees of freedom are strongly correlated, and by
spontaneously forming aggregates, they are unable to reorganize the
system. The phase space is then fragmented in a macroscopic number of
disconnected sectors (broken ergodicity), resulting in self-induced
disorder and ``thermodynamic'' anomalies, measured by the loss of a
finite configurational entropy. We discuss these results in the view
of experimental results on the kagome compounds,
SrCr$_{9p}$Ga$_{12-9p}$O$_{19}$, (H$_3$0)Fe$_3$(SO$_4$)$_2$(OH)$_6$,
Cu$_3$V$_2$O$_7$(OH)$_2$.2H$_2$O and Cu$_3$BaV$_2$O$_8$(OH)$_2$.
\end{abstract}

\pacs{PACS numbers:}
\maketitle

\section{Introduction}

Certain magnetic compounds lack conventional magnetic long-range order
but develop static order below a temperature $T_g$, with locally ``frozen''
spins. Well-known examples are spin-glasses but there are now examples
of geometrically frustrated compounds with somewhat different microscopic
properties. They are dense (one spin per site on a periodic lattice), but have a rather small ``frozen'' moment.

The physical origin of such glassy-like phases is an interesting
issue. It may be a spin-glass phase associated with weak quenched disorder\cite{Villainisg,BellierCastella,Saunders} or it may be more intrinsic to the pure
compound and its geometrical frustration. For example, structural
glasses do lack quenched disorder but are out-of-equilibrium with a
relaxation time longer than the observation time of the experiment. It
is a general idea that the frustration, by suppressing long-range
order, may lead to glassy-like phases.\cite{reviewglass,Tarjus} In the
present paper, we study the relaxation to equilibrium of the dynamics
of a simple spin model, in the presence of geometrical frustration. We find that the spin relaxation is nonexponential and
develops spontaneously two time-scales below a crossover temperature
$T_d$. The system does not slow down uniformly in space; instead, it
develops some fast-moving and slow-moving regions characterized by an
emergent length-scale (called ``dynamical heterogeneities'' in the
context of structural glasses\cite{reviewglass}). Below a second
crossover temperature $T_g$, the slow-moving spins may appear
``frozen'' on the experimental time-scale, \textit{i.e.} the system
has fallen out-of-equilibrium. In this case, the system is found to be trapped
into one of an exponential number of metastable states and some local
disorder is self-induced.

Competing local spin interactions resulting \textit{e.g.} from the
geometrical frustration of the lattice tend to suppress the magnetic
long-range order.  At low temperatures, some local correlations appear
and the system is in a collective paramagnetic regime.
The spin dynamics is different from that of a high-temperature
paramagnet: the system still has a macroscopic number of accessible
states but these states are locally constrained.  The spin dynamics is
hindered by these local constraints: single spin flips become
suppressed if they violate local arrangements and the degrees of
freedom acquire a more collective nature, which, in the present
context, are loops (or ``strings'') of spins.  The issue is whether
these cooperative excitations are efficient enough to reorganize the
system as in the liquid state (here the paramagnetic state) or if the
system is ``jammed''. Such excitations are rather ubiquitous and
appear in different contexts, \textit{e.g.} ice and ferroelectrics.\cite{Villainferro}  Stringlike
excitations have been also identified in molecular dynamics
simulations of structural glasses\cite{Kob} and were argued to indeed
play a role in the glass transition problem.\cite{Langer} Here we
study how these excitations self-organize in a simple degenerate spin
model \textit{on a lattice}, and how they permit or not, depending on
temperature, the relaxation to equilibrium. We find that, while the
motion of long loops is very efficient at high temperatures, it is too
slow at low temperatures and the residual ``rapid'' degrees of
freedom do not lead to thermodynamic equilibrium.

The magnetic materials we have in mind are highly frustrated systems
with spins on the sites of the two-dimensional kagome lattice, but
some spin-ice systems on the three-dimensional pyrochlore lattice have
a rather similar phenomenology,\cite{Gardner,Cava} and sustain similar loop excitations.\cite{MoessnerRaman} The kagome systems have a spin freezing
transition at $T_g$ but the ``frozen'' moment is rather small and the
system retains some dynamics below $T_g$.  This is the case of the
rather dense kagome bilayer SrCr$_{9p}$Ga$_{12-9p}$O$_{19}$
(SCGO),\cite{Obradors0} which was argued originally to be an
unconventional spin-glass because (i) the specific heat is in
$T^2$,\cite{RamirezC,RamirezC1} (ii) $T_g$ is weakly sensitive to the
chemical content $p$,\cite{Martinez,Ramirez} (iii) the ``frozen''
moment is small and most of the system remains
dynamical.\cite{Broholm,Lee,Mondelli,MutkaCanada,Mutka} In the kagome
hydronium jarosite,\cite{jarosite} (H$_3$0)Fe$_3$(SO$_4$)$_2$(OH)$_6$,
the $T_g$ does not depend much on the Fe coverage and compounds with
100$\%$ of Fe (as the chemical formula suggests), were
synthesized.\cite{Willspreparation} Chemical disorder is certainly not
absent, though, with possible proton disorder.\cite{Willspreparation}
Nonetheless, temperature cycles below $T_g$, were qualitatively
different from that of conventional spin-glasses, and may point to a
different nature of the phase transition.\cite{Wills,Ladieu} More
recently, two other kagome compounds were found: the
volborthite\cite{Hiroi} [Cu$_3$V$_2$O$_7$(OH)$_2$.2H$_2$O] and the
vesignieite\cite{Okamoto} [Cu$_3$BaV$_2$O$_8$(OH)$_2$]. Both have a
spin freezing transition\cite{Bert,Yoshida,Colman} with small frozen
moments.\cite{Bert,Colman,Quilliam} NMR revealed a heterogeneous state
below $T_g$: the NMR relaxation time appears to depend on the nucleus
in volborthite, with ``slow'' and ``fast'' sites found in the
lineshape.\cite{Bert,Yoshida1} In vesignieite, a partial ``loss'' of
some nuclei (partial ``wipeout'' of the intensity) is also possibly
indicative of sites with \textit{slower} magnetic
environments.\cite{Quilliam} These experiments may suggest the
presence of dynamical heterogeneities.\cite{cautiousname} These are
two-dimensional systems but a freezing transition also occurs in the
hyper-kagome gadolinium gallium garnet, Gd$_3$Ga$_5$O$_{12}$, a
three-dimensional version of the kagome lattice.\cite{Schiffer}
However, not all kagome antiferromagnets have a spin freezing
transition. Some have antiferromagnetic long-range order, such as
those of the jarosite family\cite{jarosite} (other than the hydronium
jarosite) or the oxalates.\cite{Lhotel} Others may be quantum spin
liquids, such as the herbertsmithite ZnCu$_3$(OH)$_6$Cl$_2$ that has
no phase transition\cite{Mendels} and a dynamics down to the lowest
temperatures with no clear energy scale in neutron inelastic
scattering.\cite{Helton0,deVries,Helton1} Such a broad response has
some similarities with that of
SCGO\cite{Broholm,Lee,Mondelli,MutkaCanada} or the hydronium jarosite
above the freezing temperature.\cite{Fak} This points to competitions
between different states and while it is possible to model some
antiferromagnetic phases by appropriate interactions, \textit{e.g.}
further-neighbor interactions,\cite{Harris} or Dzyaloshinskii-Moriya
interactions,\cite{Elhajal,CepasDM} the issue of the spin freezing is
delicate.

Many theoretical studies of spin freezing phenomena in the context of
the kagome antiferromagnet have been undertaken, mainly from classical
or semi-classical approaches. The role of the local collective degrees
of freedom (also called ``weathervane'' modes) was put forward,
leading to the conjecture of a spin freezing for the Heisenberg kagome
antiferromagnet.\cite{Ritchey,Aeppli} It was later argued that
distortions may help in stabilizing a ``frozen'' state, e.g. a
trimerized kagome antiferromagnet has slow dynamics on time-scales of
single spin-flips,\cite{Ferrero} (short compared with the time-scales
probed in the present study, as we shall see) or distorted kagome
lattices.\cite{Wang} It is in discrete spin models that a ``jamming''
transition was found, in the presence of additional interactions that
favor an ordered state: the dynamics becomes very slow as a
consequence of a special coarsening of the domains of the ordered
phase.\cite{Chakraborty,Castelnovo} Here we shall consider similar
discrete spins, with a different classical dynamics (not induced by
additional interactions -the equilibrium state remains paramagnetic),
but resulting from activated motion within discrete degenerate states.

The paper is organized as follows: in section \ref{Model}, we
introduce a simple degenerate spin model and the associated dynamics
within the degenerate ground states.  Section \ref{micro} gives a
heuristic motivation based on a microscopic model more appropriate to
real kagome compounds.   In section \ref{dynamics}, we present
the results of Monte Carlo simulations of the dynamics of the degenerate
model. In section \ref{Fragmentation}, we study how the phase space
gets fragmented in many metastable states and compute the configurational entropy from finite-size scaling.  We compare with experiments on kagome
compounds in section \ref{exp} and conclude in section
\ref{Conclusion}.

\section{Model}
\label{Model}

\begin{figure}[t]
\centerline{
 \psfig{file=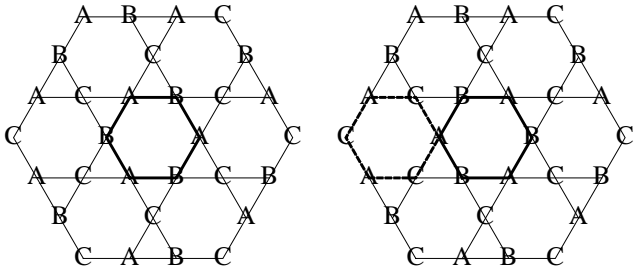,width=8.5cm,angle=-0}
}
\caption{Simplest motion compatible with the constraint: color exchange along a loop of length $L$ (\textit{e.g.} $L=6$). Note that the flip of the central loop (on the left: before the move) facilitates the motion of neighboring sites by creating a new flippable loop (on the right, after the move).}
\label{n6flip}
\end{figure}

We consider a classical three-coloring model\cite{Baxter} with spin variables $S_i=$ A, B, C (three possible colors, or spins at 120$^o$) defined on a lattice. $i$ are the bonds of the two-dimensional hexagonal lattice, or the sites of the kagome lattice (Fig.~\ref{n6flip}). There is a strict local constraint
which forces neighboring sites to be in different colors, and each
state $p$ that satisfies the constraint has energy,
\begin{equation}
E_p=0, \label{Ei=0}
\end{equation}
by definition. The number of degenerate states is macroscopic (extensive
entropy) and was calculated exactly in the thermodynamic
limit.\cite{Baxter}  As a consequence of Eq.~\ref{Ei=0}, the temperature has no effect on
the thermodynamics of the model: at equilibrium, each state $p$ has the same
probability. Yet the spin-spin
correlations averaged over the uniform ensemble are nontrivial
because of the local constraint and decay algebraically (``critical'' state).\cite{Huse}
However, the model has no dynamics and one has to specify a particular
model to study dynamical properties.

Here we consider the simplest dynamics within the degenerate states, \textit{i.e.} compatible with the constraint.  While the constraint
forbids single color changes, the simplest motion consists of
exchanging two colors along a closed loop of $L$ sites (Fig.~\ref{n6flip}).  We assume an
activation process over a barrier of energy $\kappa L$ (where $\kappa$
depends on microscopic details), with a time-scale,
\begin{equation} 
\tau_L(T) = \tau_0  \exp \left(\kappa L/T \right)
\label{tau}
\end{equation} 
where $T$ is the temperature and $\tau_0$ a microscopic time.  The
exact form (Eq.~\ref{tau}) is unessential, the important point being
that longer loops take longer time (local dynamics).  Since
the system is known to have a power-law distribution of loop
lengths\cite{Ritchey,Chakraborty,KondevHenleyPRL} (reflecting the criticality of the
thermodynamical state), we have therefore a broad distribution of
time-scales in the problem. However, the loops are strongly correlated
and the spin dynamics is nontrivial. 

It has been argued that such
constrained problems can be described at large scales by effective
gauge theories. Such examples are spin-ice systems or hard-core dimers
which can be viewed as artificial Coulomb
phases.\cite{MoessnerRaman,Henleyreview} The local constraint is solved by an auxiliary (divergence-free) gauge field and a long-wavelength
free-energy is postulated. It describes, as in standard electrostatics,
algebraic correlations at long distance.  The hydrodynamic parameters
are then extracted from the comparison with exact results (in the present case,\cite{Huse} the Baxter solution\cite{Baxter}) or
numerics. Furthermore, it also allows one to predict a relaxational dynamics
(\textit{e.g.} Langevin) and the slowest spin-spin
correlations are expected to decay as a power-law, as in dimer models.\cite{HenleydimersLangevin}

However, we also find a different ``short-time'' regime, resulting
from the microscopic model we are considering.  Indeed, the motion of
a loop reorganizes its immediate vicinity and can facilitate the
motion of a so far frozen neighbor (see Fig.~\ref{n6flip}). In this
sense, this resembles kinetically constrained models where the motion
of a local variable needs a specific configuration of its
neighbors,\cite{Palmer,Fredrickson,reviewglass} but the kinetic constraints here
result directly from the local correlations. Although the system is
fully packed with loops (each site belongs to two loops), the issue is
how the loops (and especially the small loops) self-organize.

\section{Microscopic origin of the model}
\label{micro}

We give some heuristic justifications for the model of section \ref{Model},
based on microscopic considerations. The model can indeed be viewed as an
effective model within the ground state manifold of some more general
Hamiltonian, at $T \ll JS^2$ where $J$ is defined below. We consider first a Heisenberg model,
\begin{equation}
H= J \sum_{<i,j>} \mathbf{S}_i\cdot \mathbf{S}_j 
\label{H0}
\end{equation}
where $\mathbf{S}_i$ is a quantum spin $S$ operator on site $i$ of the
kagome lattice, and $J$ an antiferromagnetic coupling between nearest
neighbor spins. We will discuss the semi-classical treatment for which
the classical states are the important starting point.

\subsection{Degenerate three-color states}

The minimization of the classical energy associated with
[Eq.~\ref{H0}] leads to many degenerate states where spins point at
120$^o$ apart on each triangle. These states are not necessarily
coplanar; however, the coplanar states have the lowest free-energy at
low $T$, a form of (partial) order-by-disorder to a ``nematic'' state.\cite{Chalker,Reimers,Zhitomirsky}
Similarly, for quantum fluctuations at order $1/S$ in spin-wave
theory, the zero-point energy is minimized by the coplanar
states.\cite{Harris,Ritchey} However, all coplanar states remain degenerate at
the harmonic level. The spins pointing at 120$^o$ in the common plane
are represented by three colors A, B, C and the three-color states
therefore form the ground state manifold of the model [Eq.~\ref{H0}].

It is a rather difficult issue to calculate the lifting of the
degeneracy due to anharmonic fluctuations. In this respect, the long-range
ordered N\'eel state with a $\sqrt{3} \times \sqrt{3}$ unit-cell plays
a special role. It was indeed argued that small-amplitude fluctuations
(albeit anharmonic, \textit{i.e.} at the next order in spin-wave theory) favor this state,\cite{Chubukov,Chan,Henleyclass} This is similar to the result of
Schwinger-boson mean-field theory,\cite{Sachdev} although this is true only at 
(small) finite $T$.\cite{Messio} From high-$T$ series expansion, the degeneracy is indeed lifted but is a small effect.\cite{Harris}

By Eq.~\ref{Ei=0}, we assume that the lifting of the degeneracy is small
compared with both the temperature and the energy barriers.

\subsection{Activation energy, quantum tunneling}

The generation of an energy barrier by fluctuations is typical of
order-by-disorder.\cite{Villain,Henleysquare} A canonical example is the
$J_1-J_2$ model on the square lattice. While two
sublattice N\'eel order-parameters can point in any direction at the
classical level, the fluctuations select the collinear
arrangements.\cite{Henleysquare} The rotation of one sublattice
order-parameter with respect to the other costs a
(fluctuation-induced) macroscopic energy.  There remains only two
degenerate states separated by an energy barrier of $O(N)$, the number of sites
(broken symmetry). In systems with a macroscopic number of degenerate
states, the situation is different because local modes
connect different degenerate states. The states are separated by
barriers of O(1) and the associated dynamics which consists of large-amplitude motion of collective spins Eq.~\ref{tau} may be relevant.\cite{Ritchey,Delft,cepasralko}
There are two different processes: the small fluctuations
about a given state of the manifold, and the large-amplitude motion
within the manifold. For continuous spins, the large-amplitude motion consists of rotating collectively the spins of a loop, out-of-plane, by an angle $\theta$ in a cone at 120$^o$, thus preserving the constraint. 
The corresponding fluctuation-induced barriers were calculated numerically and appear not to be a pure
function of the loop length as assumed in Eq.~\ref{tau}, but also depend on the
configuration.\cite{cepasralko} However, for small loops at the lowest $T$ (when the fluctuation energy is dominated by the quantum zero-point motion), $E \simeq \kappa L$ ($\kappa=0.14JS$),\cite{cepasralko} and Eq.~\ref{tau} is
justified. 

At very low temperatures, quantum tunneling through the barrier may take
place\cite{Delft} and the time-scales of Eq.~\ref{tau}
saturate.\cite{Weiss} The time-scales then depend on the barrier
shapes and the model considered.\cite{cepasralko}

In real systems, symmetry-breaking fields of spin-orbit origin may be present and provide also some energy barriers. Consider for example,
\begin{equation}
H'=H + D \sum_i (S_i^z)^2 - \sum_{i,k} E_k (\hat{\mathbf{d}}_k \cdot \mathbf{S}_i)^2,
\end{equation}
which is chosen to be compatible with the 3-coloring states: $D>0$ is
an easy-plane (xy) anisotropy and the three vectors
$\hat{\mathbf{d}}_{k}$ are directed at 120$^o$ in the kagome
plane.\cite{noteanisotropy} In the limit of strong $D$, $H'$ is
analogous to the 6-state clock model,\cite{Jose} except for the
degeneracy of the classical ground states. We note that in the
opposite limit of Ising-like XXZ anisotropy, although the system 
orders ferromagnetically, there is a slow persistent dynamics of creation of loops.\cite{Miyashita} We will restrict the discussion to $D>0$
and $E_k=0$ in the following.

When the anisotropy is small (which is generally the case of 
intermetallic magnetic ions), rotating the
spins of a loop continuously by $\theta$, defines a classical energy barrier, $\kappa L \sin^2 \theta$ with $\kappa= 3DS^2/2$.

When the anisotropy is strong (possibly more appropriate to rare-earth
compounds), it is too costly to rotate all components
out-of-plane. The lowest-energy excitations consist of violating the
constraint by nucleating defects.  The simplest effective process for
the spins to move in the constrained manifold is to create two defects
along a loop. This costs twice the exchange energy but the defects are
then free to move along the loop (deconfinement) and leave behind them
a string of exchanged colors.\cite{Castelnovo} When the defects
annihilate, the loop has flipped. The time-scale of this process is
given by Eq.~\ref{tau}, with $\kappa \sim JS^2$.\cite{Castelnovo} This
is the important effective process in spin-ice systems in general, and
the nonequilibrium dynamics of defects has been directly studied
recently.\cite{Moessner,Cugliandolo}

Note that by using the discrete model, we 
intend to describe only the slow collective degrees of freedom.
The rapid motion about the ``equilibrium'' state (spin-waves) is present in the continuous spin model but is integrated out in the discrete model (in the barrier), at the first order of spin-wave theory.\cite{cepasralko}

\section{Stochastic spin dynamics}
\label{dynamics}

The system evolves in the classical degenerate manifold by the motion
of closed loops, described by a local stochastic activated process
given by Eq.~\ref{tau}.  

We have studied the spin (color) dynamics by classical Monte Carlo
simulations.  Such Monte Carlo simulations have been used to study the
equilibrium state of constrained or loop
models,\cite{Stillinger,Melko,Jaubert} and also in the present
context.\cite{Huse,Chakraborty,Castelnovo} In these simulations, the
updates were accepted following the Metropolis algorithm, and irrespective
of the length of the loop.  Here, the aim is not to probe the
equilibrium state (which is known) but to study how the spin dynamics
slows down when longer loops have to pass higher energy barriers,
which take more time. The issue is rather to study the relaxation to
equilibrium.

The algorithm is similar to that used earlier: (a) we choose a single
site at random, (b) we choose a neighbor of this site at random (this
defines two colors, hence one of the three types of loop A-B, A-C, or
B-C) (c) we search among its four neighbors the site with the same
color as the original site (but distinct from it) and we iterate until
a closed loop is formed (this is guaranteed by the periodic boundary
conditions).  Contrary to previous studies, however, the colors are
exchanged along the two-color loop (the loop is ``flipped'') according
to the probability to cross the barrier, $1/\tau_L$.  This amounts to
choose in the Metropolis acceptation rate a microscopic rate which
depends on the degree of freedom that moves. The cluster sizes are
$N=3L^2$, $L$ is the linear size (up to $L=144$) and a Monte Carlo
sweep (MCS) corresponds to $N$ attempted updates.

We have computed the
autocorrelation function
\begin{eqnarray}
C(t) &=& \left\langle \frac{1}{N} \sum_{i=1}^N  \mathbf{S}_i(t).\mathbf{S}_i(0) \right\rangle 
\end{eqnarray}
where $\left\langle \cdots \right \rangle$ is an average over initial
states (10$^3$ in Fig.~\ref{Sloct}, and up to 10$^4$ for better statistics).  Here, because we have a three-color model,
$\mathbf{S}_i(t).\mathbf{S}_i(0)=1$ for parallel spins (same color)
and $-1/2$ for spins at 120$^o$ (different colors). By definition,
$C(0)=1$ and if the state at time $t$ is decorrelated from that at
$t=0$, each spin is in one of the three possible colors with
probability $1/3$ and $C(t)=0$ ($C(t)$ measures how long the system
keeps memory of its initial state).  To accelerate the simulations, we
rescale Eq.~\ref{tau} by $\tau_{\beta} \equiv \tau_6(T)$ so that the
shortest loops (hexagons) flip at each attempt. In the following, the
MCS are in units of $\tau_{\beta}$ and $T$ in units of $\kappa$.  The
Fourier transform of $C(t)$, $C(\omega)$, is the local spin
susceptibility, as measured by experimental probes, for instance
neutron inelastic scattering (cross-section integrated over all
wave-vectors), NMR or $\mu$SR on different time-scales.

\subsection{Summary of the results}

\begin{figure}[h]
\psfig{file=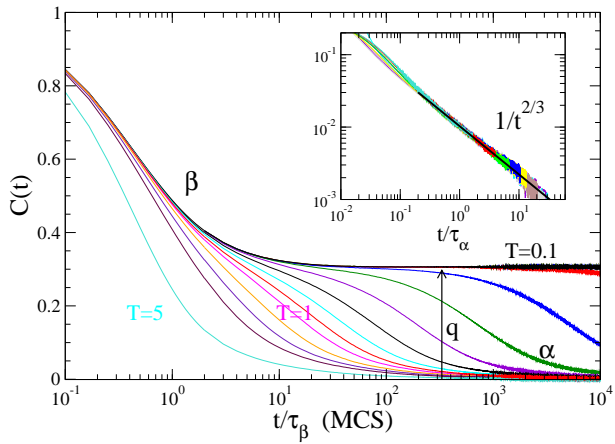,width=7cm,angle=-90} 
\caption{Spin autocorrelation as a function of time (Monte Carlo sweeps) with decreasing $T$ (from left to right), $C(0)=1$ ($L=144$). Inset: long-time tail (rescaled), $1/t^{2/3}$ (solid line), as described by a height model.}
\label{Sloct}
\end{figure}

The autocorrelation is given in Fig.~\ref{Sloct} for different
temperatures.  The relaxation of the system occurs on a time-scale
$\tau_{\alpha}$ and follows a power-law decay, $t^{-2/3}$ (inset of
Fig.~\ref{Sloct}), which is well described by a long wavelength field
theory, as we shall see. 

Below a crossover temperature $T_d$, the
spin dynamics develops two distinct time-scales, $\tau_{\alpha}$ and $\tau_{\beta}$ ($\tau_{\alpha}$ and $\tau_{\beta}$ are the notations in
supercooled liquids for the long and short relaxation times): the autocorrelation decreases first into a plateau (quasi-stationary state) and then relaxes to equilibrium.  At short times $\sim \tau_{\beta}$,
the relaxation is approximately a stretched
exponential $C(t) \approx \exp(-t^{\beta})$ ($\beta \approx 0.63$). While the dynamics is spatially homogeneous above $T_d$, it becomes heterogeneous below $T_d$ with slow and fast regions.

\begin{figure}[h]
\psfig{file=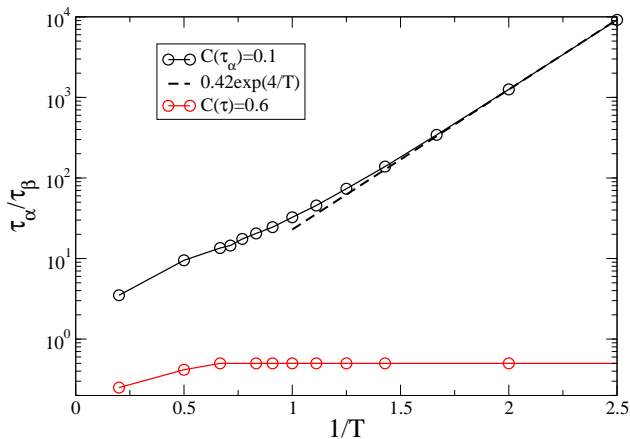,width=6cm,angle=-90} 
\caption{Two relaxation time-scales.}
\label{twotau}
\end{figure}

\subsection{Long-time relaxation}

We define the relaxation time of the system, $\tau_{\alpha}$, by
\textit{e.g.} $C(\tau_{\alpha})=0.1$ (the value chosen has no consequence as long as it is small enough). In Fig.~\ref{twotau}, we give
$\tau_{\alpha}/\tau_{\beta}$ as a function of temperature. This ratio
becomes much larger than one in the limit of low-$T$,
$\tau_{\alpha}/\tau_{\beta} \approx 0.42\exp(4/T)$ so that
$\tau_{\alpha} \sim \tau_{10}(T)$ is controlled by the second shortest
loops (of length 10). For comparison, the time that characterizes the
initial decay of $C(t)$, defined by $C(\tau)=0.6$, is of order
$\tau_{\beta} \equiv \tau_{6}(T)$ (Fig.~\ref{twotau}), \textit{i.e.}
controlled by the shortest loops.  Such definitions and spontaneous
generations of two time-scales appeared in a different spin model
where the frustration is played by long-range interactions which
fragment the system into domains.\cite{TarjusViot}

By rescaling all the curves by $\tau_{\alpha}$, we find that the decay
at long times is a power-law,
\begin{equation}
C(t) \sim 1/t^{1-\alpha},  
\label{1t}
\end{equation} with $\alpha \approx 0.33$ (see the inset of Fig.~\ref{Sloct}). Since $\alpha>0$, the integrated relaxation time $\int_0^{\infty} C(t) dt$ diverges, and, at small frequencies, the Fourier transform diverges like $\omega^{-\alpha}$ (we do not discuss here some natural cutoffs provided by \textit{e.g.} defects at finite temperatures).

 The long-time regime reflects the criticality of the equilibrium
 state and is well described by a free vector-field model. The model
 is obtained by a mapping of the color variables onto an auxiliary two-component
 height field $\vec{\varphi}$ defined at the centers of the
 hexagons.\cite{Huse,Read,Kondev} The construction is as follows: the
 height vector $\vec{\varphi}$ picks up a $\hat{e}_i$ vector each time
 it crosses a $i=$A,B,C color with the condition
 $\hat{e}_A+\hat{e}_B+\hat{e}_C=0$. In such a way, the local
 constraint is automatically satisfied.  One assumes that the
 free-energy (of purely entropic origin) reads
\begin{equation}
F/T=   \frac{1}{2} K \int d^2 \mathbf{x} (\nabla \vec{\varphi})^2 
\label{F}
\end{equation}
where $\vec{\varphi}$ is the coarse-grained height field. The stiffness $K=2\pi/3$ is chosen such as to reproduce the exact critical exponent of the spin-spin algebraic correlations,
$\eta=4/3$.\cite{Huse,Read,Kondev} Eq.~(\ref{F}) describes a classical\cite{notedeconfinement}
interface in two spatial dimensions. Similarly to dimer models,\cite{HenleydimersLangevin} the classical fluctuations of the interface can be described by Langevin
equations, 
\begin{equation}
\frac{\partial \vec{\varphi}}{\partial t}= D \nabla^2 \vec{\varphi} +\vec{\eta} (\mathbf{x},t)
\label{Langevin}
\end{equation}
where $\vec{\eta}(\mathbf{x},t)$ is a two-dimensional white noise, $\langle \vec{\eta}(\mathbf{x},t).\vec{\eta}(\mathbf{x}',t') \rangle=T \delta(\mathbf{x}-\mathbf{x}')\delta(t-t')$. Eq.~\ref{Langevin} describes a simple diffusion of the height of the interface. The mapping to the slowest spin fluctuations, $m_s(\mathbf{x},t)=e^{iQ.\vec{\varphi}(\mathbf{x},t)}$, $|Q|=4\pi/\sqrt{3}$,\cite{Huse,Read,Kondev} gives the spin correlations at long times and long distance,
\begin{equation}
C(\mathbf{x},t)=\langle m_s(\mathbf{x},t)m_s(0,0) \rangle \sim \frac{1}{t^{1-\alpha}} f\left( \frac{| \mathbf{x}|}{t^{1/z}} \right)
\label{talpha}
\end{equation}
with $1-\alpha=\eta/z$, $z=2$ (from Eq.~\ref{Langevin}) and $f(0)=1$. We therefore obtain
$\alpha=1/3$, in good agreement with the
$1/t^{1-\alpha}=1/t^{2/3}$ found numerically (see inset of Fig.~\ref{Sloct}).
The approach also explains
that the exponent does not vary with $T$ because the underlying critical phase is independent of $T$, by definition. 

Eq.~\ref{talpha} characterizes the spin fluctuations at long times (by definition
of the coarse-grained free-energy). At short times, however, 
corrections to Eq.~\ref{Langevin} are important and lead to a
different dynamics, as we now show.

\subsection{Short-time and plateau below $T_d$}

Below a crossover temperature $T_d \approx 1$, a shoulder develops in
$C(t)$ and the relaxation time $\tau_{\alpha}$ starts to differ
from  $\tau_{\beta}$ which characterizes the initial decay.  $C(t)$ develops a plateau which becomes more and
more stable when $T$ is further lowered.
The limiting value of the plateau is (see Fig.~\ref{Sloct}),
\begin{equation}
q \equiv \frac{1}{N} \sum_{i=1}^N \langle \mathbf{S}_i \rangle^2 \approx 0.31.
\label{qnn}
\end{equation}
It gives the averaged frozen moment on time-scales shorter than $\tau_{\alpha}$, which we note $\langle
\mathbf{S}_i \rangle \approx 0.56$.  On these
time-scales, only the hexagons have dynamics: all
other loops are blocked until $\tau_{\alpha} \approx \tau_{10}(T)$ at
which a loop of length 10 may flip, and the system leaves the plateau
and returns to equilibrium.

When the relaxation time of the system becomes longer than the experimental time, $\tau_{\alpha} \approx \tau_{exp}$, the system is out-of-equilibrium.  This occurs at the glassy-like crossover temperature, $T_g<T_d$ (which depends on the typical time-scale of the experiment). From 
the estimation of $\tau_{\alpha}$, we have $T_g =10/\ln(\tau_{exp}/(0.42\tau_0)) \approx 0.3$ for $\tau_{exp}=10^3$s and $\tau_0 \sim 10^{-12}$s. For $T<T_g$ the system is trapped into the plateau. Once all fast processes have occurred (\textit{i.e.} after $\tau_{\beta}$) the system is in a quasi-stationary state with frozen moment squared $q$ (we reserve the term ``Edwards-Anderson
order-parameter'' to true equilibrium phase transition).

\begin{figure}[h]
\centerline{
 \psfig{file=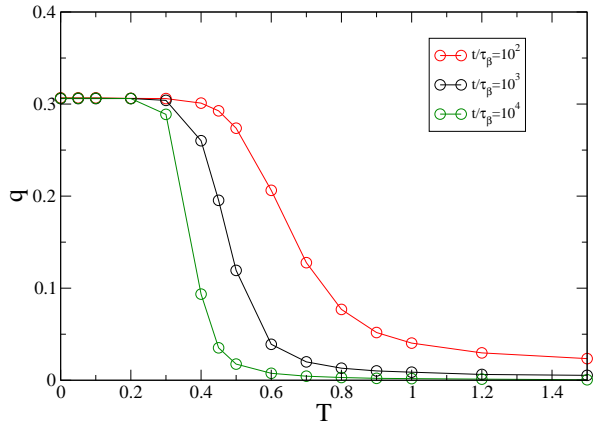,width=7cm,angle=-90}
}
\caption{Frozen moment squared as a function of temperature and observation time, Eq.~\ref{qnn}. $L=144$.}
\label{q}
\end{figure}

We can furthermore calculate $q$ as a function of
$T$. It is related to the susceptibility by
\begin{equation}
\chi \equiv \frac{\langle \mathbf{S}_i^2 \rangle - \langle \mathbf {S}_i\rangle^2}{T} = \frac{1-q(T,t)}{T}
\end{equation}
The frozen fraction depends logarithmically on time
below $T_g$ (see Fig.~\ref{q}), so that $\chi$ has a cusp at $T_g$
between a high-$T$ paramagnetic susceptibility $\chi=1/T$ and a low-$T$
time-dependent susceptibility.

\vspace{.4cm}

\begin{figure}[h]
\psfig{file=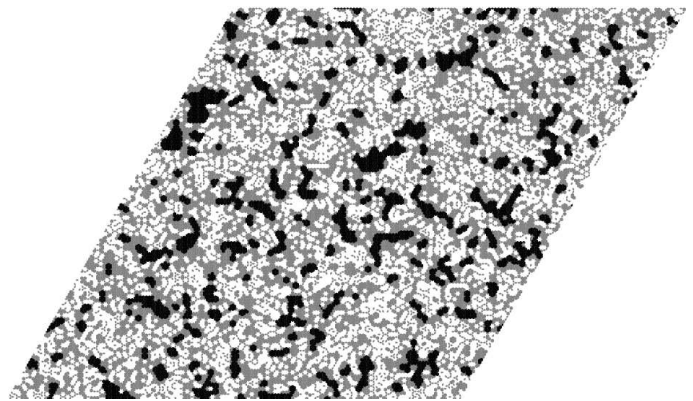,width=9cm,angle=-0}
\caption{Real space picture of the autocorrelation,
  $C_i(t)=\mathbf{S}_i(t).\mathbf{S}_i(0)$, at time $t=10^3$ and $T=0.1$. Black: frozen sites. White ($C_i(t)=-1/2$), in gray the sites which have moved between $0$ and $t$ but have returned to their initial state ($C_i(t)=1$).}
\label{overlap}
\end{figure}

The existence of a frozen moment on average is both the consequence of
frozen regions (which are purely static) and dynamical regions with a
finite moment on average (because of a recurrent behavior).  In
Fig.~\ref{overlap}, we show the autocorrelation
$C_i(t)=\mathbf{S}_i(t).\mathbf{S}_i(0)$ on each site, at intermediate
times in the quasi-stationary state ($-1/2$ is white, $1$ is black if it has
never moved between 0 and $t$ or gray otherwise). While most sites
have dynamics (white and gray), there is a fraction of frozen sites
(in black).  The averaged fraction of frozen sites is $N_f=0.121N$,
and the probability distribution function is found to be gaussian (as a
consequence, Fig.~\ref{overlap} is typical of what happens at low
$T$). The existence of 12.1$\%$ of frozen sites explains only part of
the averaged frozen moment, $q=31\%$.  In addition, other (dynamical)
sites contribute. This is because the frozen clusters provide boundary
conditions for the neighboring sites and the constraint propagates
between clusters. For instance, the spins on the outer side of the
cluster boundary can take only two of the three possible states, the
third possibility being frozen inside the cluster. They have hence
stronger probabilities to return to the original value.  In
Fig.~\ref{overlap}, we indeed see, extending between the frozen
clusters, large dynamical regions where the spins are in their
original state (gray).  These constrained regions contribute to almost
two thirds of the averaged frozen moment.

\vspace{.4cm}
\begin{figure}[h]
\centerline{
 \psfig{file=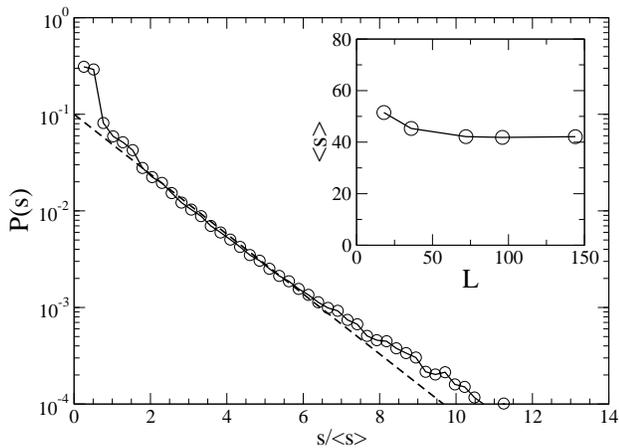,width=6.5cm,angle=-90}}
\caption{Distribution of the sizes of the frozen clusters. The average is $\langle s \rangle=42$ sites (inset: finite-size effect) or the emergent length scale is $\langle s \rangle^{1/2}$. The dashed line is a guide to the eye (exponential).}
\label{histcluster}
\end{figure}
\vspace{.4cm}
\begin{figure}[h]
\centerline{
 \psfig{file=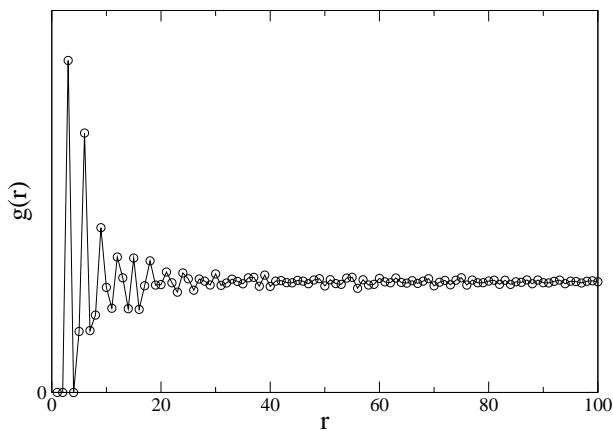,width=7cm,angle=-90}}
\caption{The radial distribution function of active degrees of freedom: probability to have a flippable hexagon at distance $r$ from a given flippable hexagon at 0 (normalized by the number of hexagonal sites) averaged over the uniform ensemble. While the nearest neighbor position is not compatible with the constraint, the first peak corresponds to an attraction of next nearest neighbor hexagons.}
\label{structurefactorloop}
\end{figure}

Furthermore, it is seen in Fig.~\ref{overlap} that frozen sites form
clusters randomly distributed over the system.  The number of spins in
a cluster is distributed according to Fig.~\ref{histcluster}. The
average is $\langle s \rangle=42$ sites (and is size independent for
$L \gtrsim 72$, see inset of Fig.~\ref{histcluster}), thus defining an emergent
length scale $\langle s \rangle^{1/2}=6.5$ intersite spacings.  The
picture of the frozen phase is that of ``jammed'' clusters of nanoscopic
scale $\langle s \rangle^{1/2}$ occupying 12.1$\%$ of the sites.

\begin{figure}[h]
 \psfig{file=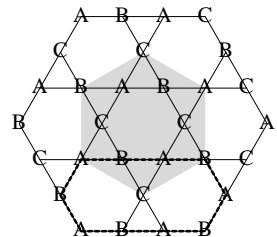,width=3.5cm,angle=-0}
\caption{``Jammed'' cluster (the smallest one, in gray): no 6-loop can unjam any of its 12 sites. The shortest ``unjamming'' loop (of length 10) is shown (dashed line).}
\label{jam}
\end{figure}

What is the origin of the ``jamming''? First, ``jammed'' clusters do
not contain flippable hexagons (by definition) but are, of course,
criss-crossed by longer loops which are blocked at the temperatures
considered. This implies that a typical three-coloring state must have
a low-enough density of flippable hexagons. On the kagome lattice the
density of flippable hexagons (averaged over the uniform ensemble) is
0.22, so that forming a large cluster of $\langle s \rangle=$42 sites
on average is unlikely in absence of correlations. In
Fig.~\ref{structurefactorloop}, we give the correlations $g(r)$
(radial distribution function) in the positions of the flippable hexagons
of the same type.\cite{Gleb} We find indeed a strong attraction: the
neighboring hexagons cannot be occupied by the same type of loop (it
is incompatible with the constraint) but the second neighbor positions
are highly favored (attraction). There is a high probability to have a
flippable hexagon if the (second) neighbor is a flippable
hexagon. This attraction creates aggregates and voids, opening the way
to regions free of flippable hexagons. The system can therefore be
viewed as a microscopic phase separation of active and inactive
regions, the active regions having flippable hexagons, the inactive
regions having longer loops. Recall that the degenerate model can be
seen as being at the boundary of a phase transition in parameter
space,\cite{Baxter} in particular between active and inactive phases,
having respectively short and long loops.\cite{lr} This is a necessary
but not a sufficient condition for the region to be ``jammed'' because
the number of flippable hexagons is not conserved by the dynamics and
they ``move'' on the lattice (see Fig.~\ref{n6flip}).  The frozen
clusters correspond to special configurations and regions inaccessible
to flippable hexagons. For example, a frozen cluster of 12 sites is
shown in Fig.~\ref{jam}: each hexagon on the border has the three
possible colors A, B, C, thus making it impossible to create a
flippable configuration. One can have clusters of arbitrary size (see
Fig.~\ref{overlap}) or walls that prevent flippable hexagons to
diffuse in different regions of the sample.

However, a loop of length 10 (shown by a dashed line in
Fig.~\ref{jam}) will unjam the configuration, and the cluster shown
will be annihilated.  The way the relaxation takes place at longer
times is via the dynamics of creation and annihilation of ``frozen''
clusters on time-scale $\tau_{10}(T)$. For $T<T_d$, there is a
separation of time-scales between the ``rapid'' hexagon motion $\sim
\tau_6(T)$ and the longer creation/annihilation of frozen clusters
$\sim \tau_{10}(T)$.

\subsection{Dynamical heterogeneities $T<T_d$}

We now consider some dynamical local quantities. 
Following studies of standard glasses,\cite{reviewglass} we define a local mobility field $K_i(0,t)$ which
measures how many times the site $i$ has changed color during the time
interval between 0 and $t$. It is linear in $t$ for large $t$ so that one can define a local frequency $f_i=K_i(0,t)/t$.
Frozen sites have $f_i=0$ while dynamical sites have $f_i>0$.

\begin{figure}[h]
\psfig{file=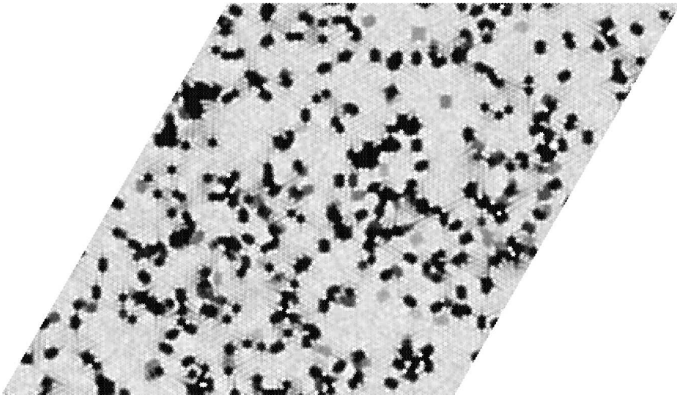,width=9cm,angle=-0}
\caption{Dynamical heterogeneities in space. The gray scale is proportional to the local frequency $f_i$ of the site from black (frozen) to white (high frequency).}
\label{realspace}
\end{figure}
 
\begin{figure}[h]
\centerline{
 \psfig{file=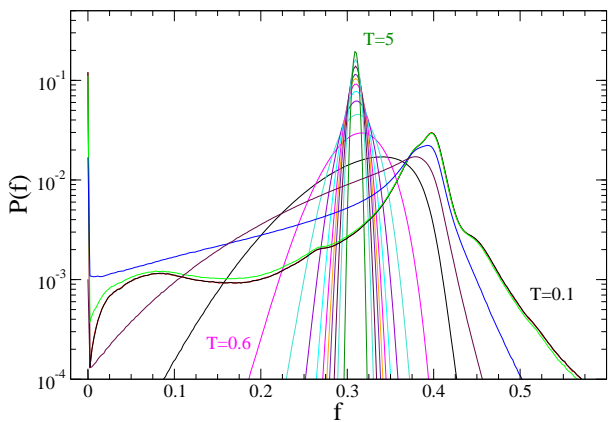,width=7cm,angle=-90}
}
\caption{Non-uniform slowing down of the dynamics by lowering the temperature. From $T>T_d$, homogeneous (gaussian) distribution of local frequencies to $T<T_d$, heterogeneous (skewed) distribution and at $T<T_g$ a frozen fraction appears. $t=10^4$ and $L=144$.} 
\label{histfreq}
\end{figure}
The real space picture of $f_i$ at a given time is given in
Fig.~\ref{realspace} from black (frozen sites) to white (fast sites):
we see the variations of the local dynamics across the system and
some clusters of slow frequencies, \textit{i.e.} a form of dynamical
heterogeneity. We plot the corresponding histogram of frequencies in
Fig.~\ref{histfreq} at various temperatures. At high temperature, the
distribution is homogeneous (gaussian).
At lower temperatures, the dynamics slows down, and the distribution broadens and becomes
asymmetric (non zero third moment or skewness). Eventually at $T<T_g$, a frozen
fraction appears and the distribution becomes continuous between two
typical peaks,
\begin{equation}
 P(f)= \frac{N_f}{N}\delta(f) +A(f)
\end{equation}
where $A(f)$ is a smooth broad function.
One can describe this evolution as a crossover between a homogeneous high-temperature phase with a single type of dynamical site and a low-temperature phase with many inequivalent dynamical sites. It can be described in terms of large-deviation functions and a ``free-energy'' can be defined.\cite{Garrahan}

\section{Fragmentation of the phase-space}
\label{Fragmentation}

We show that the phase space is fragmented into a $e^{NS_c}$ number of
sectors for $T<T_g$, separated by barriers of $O(1)$.  For this, we
directly enumerate all the states of small clusters and analyze how
the system evolves in the phase-space as a function of temperature.
This allows us to describe the landscape of energy barriers separating
states and basins, \textit{i.e.} a hierarchical organization of the states
(nonfractal here).

Let $P_p(t)$ be the probability of the system to be in a configuration
$p=1,\dots,N_C$, $N_C$ the total number of states which we have
numerically enumerated on small clusters with periodic boundary
conditions ($N=27,36,81,108$). We have found $N_C= 6.4 \times 1.122^N$ (dashed line in
Fig.~\ref{nstates}), slightly smaller than the exact result in the
thermodynamic limit $1.135^N$.\cite{Baxter}  

The master equation governing the
dynamical evolution of $\mathbf{P}(t)=(P_1(t),\dots,P_{N_C}(t))$,
\begin{equation}
\frac{ \partial \mathbf{P}}{\partial t} = \mathbf{w} \cdot \mathbf{P}
\label{master}
\end{equation}
involves a matrix $\mathbf{w}$ which contains the transition rates from a configuration $p$ to $p'$. The only allowed transitions are single flips of loops of length $L$, $w_{p \rightarrow p'}=-1/\tau_L(T)$ where $\tau_L(T)$ is given by Eq.~\ref{tau}. Here from detailed balance, we have $w_{p \rightarrow p'}=w_{p' \rightarrow p}$ ($E_p=0$ for all states) and $w_{p \rightarrow p}=\sum_{p' \neq p} w_{p \rightarrow p'}$ ensures the conservation of the probability,
$\sum_p P_p(t)=1$. 
All states satisfying
\begin{equation}
\mathbf{w} \cdot \mathbf{P}=0
\end{equation}
are stationary, such as, in particular, the equilibrium uniform
distribution $P_p(t)=1/N_C$. $\mathbf{w}$ may have more than one
zero eigenvalue and the additional stationary states 
prevent the system from exploring the phase space (broken ergodicity). Examples are systems with a broken-symmetry, the phase space
of which has a finite number of disconnected sectors in the
thermodynamic limit. In each sector the Gibbs distribution is
stationary assuring as many zero eigenvalues as the number of sectors
or broken symmetries.  By contrast, in a glassy-like phase, 
 the number of eigenvalues satisfying $\epsilon \ll
1/\tau_{exp}$ ($\tau_{exp}$ is the experimental observation time)
scales like $e^{NS_c}$: there is a finite configurational entropy $S_c$. In other words, a
macroscopic number of states, thus differing at the microscopic scale,
never relax on the observation time-scale.

In the present model, $\mathbf{w}$ has a finite hierarchical
structure.  Here it is a consequence of a microscopic model and is \textit{not
assumed} from the beginning as in hierarchically constrained
models.\cite{Palmer,Ogielski,Melin} Contrary to these examples (or
spin glasses\cite{Mezard}), however, we find only four levels of
hierarchy:  the phase space is split into a few ``Kempe'' classes,\cite{Mohar,cepasralko} which are split into $\sim N$ topological sectors and then in $e^{NS_c}$ trapping sectors (see Fig.~\ref{nstates} for a graphical illustration of this hierarchy in the phase space).

\subsection{Infinite barriers}

The dynamics of loops of all sizes is known to be nonergodic
on the kagome lattice.\cite{Huse,Mohar} It means that moving all loops is not sufficient to go from a given state to any other state in the
phase space. $\mathbf{w}$ split in ``Kempe'' classes,\cite{Mohar,cepasralko}
the number of which is in general unknown.\cite{Mohar}

Since it is therefore impossible to enumerate all states by moving
loops iteratively, we have allowed to introduce defects that violate
the 3-colored constraint. To control the density of defects, we have
introduced an energy penalty, \textit{i.e.} the antiferromagnetic three-state
Potts model. By cooling the system at low temperatures in a Monte
Carlo simulation, one generates three-coloring ground states that are
in different ``Kempe'' sectors (and the sectors themselves by
switching on the loop dynamics).  For $N=108$, we find four sectors, a
large one with $89\%$ of all states and three smaller ones, all separated
by infinite barriers for the loop model.

\begin{figure}[h]
\centerline{
\psfig{file=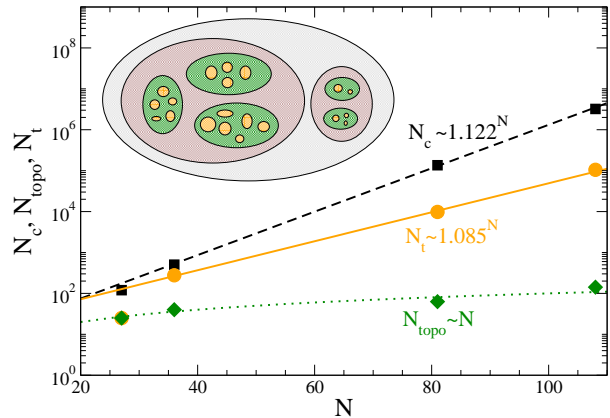,width=7cm,angle=-90}
}
\caption{Hierarchical structure of the phase space and exponential number of disconnected sectors $N_t$  at low temperatures (solid circles). $N_t$ corresponds also the number of zero eigenvalues of the matrix $\mathbf{w}$. The total number of states $N_c$ (squares), topological sectors $N_{topo}$ (diamond) result from the complete enumeration of states on clusters of size $N=27,36,81,108$.}
\label{nstates}
\end{figure}

Within each Kempe sector, the three-coloring states can be characterized
by topological numbers. They are defined by counting the number of
colors along nonlocal horizontal and vertical
cuts.\cite{Castelnovotopo} There are six such numbers $w_i^{x,y}$
($i=1,2,3$), which may take any integer value from 0 to $L$ with the
constraint $\sum_{i=1}^3 w_i^{x,y}=L$, so four of them are
independent. This gives at most $N^2$ sectors, but since some
combinations are not allowed, the number is of order $N$
(Fig.~\ref{nstates}). The dynamics of local loops conserves these
 numbers so that each Kempe sector is divided into $N$
topological sectors. Only winding loops of length  $L$ or $L^2$ (the longest
loop takes all two color sites and has length $2N/3$) may change them. In fact, the averaged length of the winding loops scales like $L^{3/2}$.\cite{KondevHenleyPRL,Chakraborty} The topological
sectors are therefore separated by barriers growing with the system
size like $L^{3/2}$, defining infinite barriers in the thermodynamic limit and broken ergodicity sectors. This is analogous to the ``jamming'' transition induced by additional forces: the favored ordered state needs rearrangements of infinite loops in order to equilibrate.\cite{Chakraborty,Castelnovo} Here we recall that the phase space is in general broken into $\sim N$ sectors (which we have explicitly constructed), labelled by quantities conserved by the local dynamics.\cite{Castelnovotopo}

\subsection{Fragmentation in $e^{NS_c}$ sectors}

For $T < T_g$, the dynamical matrix $\mathbf{w}$ split further into new smaller sectors which we have constructed for different system sizes. We find that the phase space is split into 1.085$^N$
independent trapping sectors (Fig.~\ref{nstates}).  The spin dynamics
has a fast equilibration within a sector characterized by the motion of
6-loops on time-scale $\tau_{\beta}=\tau_6(T)$ and the motion between
sectors occurs on time-scale, $\tau_{\alpha} \sim \tau_{10}(T)$, which is
frozen below $T_g$ by definition.  Above $T_g$, the system equilibrates within a
topological sector.

The number of sectors defines a
finite averaged configurational entropy per site $S_c=\ln 1.085=0.082$, which is approximately two thirds of the full entropy $S_{eq}= \ln 1.122=0.115$.  Upon reducing the temperature, the system goes from an equilibrated state with the full entropy $S_{eq}$ (the number of topological sectors is sub-extensive) to a metastable state below $T_g$ where it looses the configurational entropy:
\begin{equation}
\Delta S = S_c = 0.082= 0.7 S_{eq}
\end{equation}
The configurational entropy reflects in phase space
the entropy of the microscopic arrangements of the frozen clusters (section \ref{dynamics}).  A crude comparison consists
of distributing $N_f/\langle s \rangle$ disks on the lattice ($N_f/\langle s \rangle$ is the number of frozen clusters of average size  $\langle s
\rangle=42$, we note $x$ the density),
with entropy $S/N \sim (-x\ln x-(1-x)\ln (1-x))/\langle s
\rangle=0.009$ ($x=12\%$). This is too small though by an order of magnitude
compared with $S_c$.

For $T_g<T<T_d$, one can define coarse-grained states by eliminating
the fast dynamics into an entropy. While, on average, each sector
contains $(1.122/1.085)^N=1.034^N$ states (thus defining the averaged
entropy $S_2=0.034N$), we find a broad distribution of sector sizes from $s=1$ (a
single state) to a large sector $s \lesssim N_C$.  However, we believe
that this is a finite-size effect. Indeed the probability to fall into
a sector of size $s$ is found to be roughly constant at small $s$ and
increases for larger sectors. In contrast, for a Monte Carlo sampling
of states as done in section \ref{dynamics}, the frozen fraction distribution is homogeneous (gaussian)
for $L \gtrsim 18$, while for $L \lesssim 18$, a large portion of
states has no frozen fraction at all.  As a consequence, the
distribution of entropies is certainly more homogeneous for large
system size.

In summary, we find that the phase space has hierarchical levels: it has sectors
characterized by conserved quantities and separated by infinite
barriers (broken ergodicity) and sectors or traps separated by finite
barriers. The number of topologicals sectors is of order $N$
(nonextensive entropy), and there is no essential difference between
them at the microscopic or mesoscopic scale: a local measurement cannot distinguish between two different sectors.  On the other hand, the
number of traps is of order $e^{NS_c}$ (finite configurational
entropy). Therefore the system looses a finite entropy at $T_g$ and a
local disorder is self-induced: a local measurement can distinguish between two metastable states (for instance, if there is or not a frozen cluster). In this sense, $T_g$ can be called a glassy crossover temperature. By opposition, the jamming transition found in Refs.~\onlinecite{Chakraborty,Castelnovo} corresponds to broken ergodicity associated with a sub-extensive entropy (no self-induced disorder).

\section{Discussion of experiments}
\label{exp}

We now discuss the kagome compounds that have a freezing
transition. We argue that the freezing temperature $T_g$ is governed
by the energy scale of the barriers and when possible, we identify the
possible mechanisms we have discussed in section \ref{micro}: the
barriers are either dynamically generated by the rapid spin-wave
motion or generated by anisotropies, depending on specific materials.
We also compare the strength of the ``frozen'' moment to the
experiments available and the dynamics of the system.  Note that the
present dynamics of loops is classical (if a quantum coherence is
maintained, the system was predicted to order\cite{cepasralko}).  Some
quantum fluctuations are therefore neglected here, but may turn out to
be important, especially for the copper oxides discussed below
($S=1/2$), if the anisotropy is small enough.\cite{CepasDM}

\subsection{SrCr$_{9p}$Ga$_{12-9p}$O$_{19}$ (SCGO)}

In SCGO, a phase transition occurs at $T_g \sim 3.5-7$K, depending
weakly on the Cr$^{3+}$ ($S=3/2$) coverage $p$.\cite{RamirezC,RamirezC1,Martinez,Ramirez} $T_g$ depends also on the
experiment: $T_g \sim 3.5$K by squid, $5.2$K by neutron
scattering for the same compound.\cite{MutkaCanada}

What could be the appropriate microscopic model? The Cr$^{3+}$ ions
have no orbital moment ($L=0$) and the spin anisotropy is expected to
be small. From EPR indeed, $DS^2 \sim 0.2$K.\cite{Ohta} By constrast,
the measurements of the spin susceptibility on single crystals showed
a large anisotropy disappearing when increasing the
temperature.\cite{SchifferSCGO} This was therefore attributed to the
spontaneous breaking of the rotation symmetry by a nematic order
(coplanarity), and not a real anisotropy of the
model.\cite{SchifferSCGO} Similarly, the 8K barrier obtained by
$\mu$SR for $p\rightarrow 0$ which was originally interpreted as a
large single-ion anisotropy,\cite{Keren} is in fact absent if one uses
a different fit of the data.\cite{Bono} On the other hand, for
$p\rightarrow 1$, energy barriers of $\sim$30K were
obtained.\cite{Keren,Bono} Since they are two orders of magnitude
larger than the spin anisotropy, they are more likely to be induced by
the fluctuations.  With $E=\kappa L=30$K and $L=6$, we have $T_g = 0.3
\kappa=1.5$K.  On the other hand, if we use $\kappa=0.14JS$ (section
\ref{micro}) and $J \sim 50$K from the spin susceptibility, we find
$T_g =0.04 JS \sim 3$K.  Both estimates are in fair agreement with the
experimental result. The model does not predict, however, a
thermodynamic transition while, experimentally, this has been a
disputed point, especially regarding the sharpness of the nonlinear
susceptibility $\chi_3$.\cite{RamirezC,Obradors} We also note that the
``thermodynamic'' anomalies we have mentionned at $T_g$ are not only
rounded but also the entropy change $\Delta S=0.082N$ is small
compared with the full entropy $N \ln(2S+1)$ of continuous
spins. Yet this makes a definite prediction for the entropy change.

Furthermore, the frozen moment measured in neutron elastic scattering
is small, $\langle S_i \rangle^2 \sim 0.12-0.24$ of the maximum moment
(depending on the Cr coverage) and most of the signal is in the
inelastic channel.\cite{Broholm,Lee,MutkaCanada} In the experimental
setup of Ref.~\onlinecite{Broholm}, the inelastic channel starts above
the neutron energy resolution of 0.2 meV, giving in that case a
lifetime of the frozen moment longer than $\sim 20$ps. Neutron
spin-echo showed that the moment is still frozen on the nanosecond
time-scale at 1.5 K.\cite{Mutka} However, no static moment was originally
observed in $\mu$SR,\cite{Uemura} but a weak static component may not be
excluded.\cite{Bono} Similarly, in Ga NMR, the wipeout of the signal
shows a dynamics that has slowed down but is still
persistent.\cite{Limot} However, in both cases the muon or the Ga
nuclei probe many sites and may see primarily the dynamical sites.

In the model developed above, the system remains dynamical below
$T_g$. The system has flippable hexagons on time-scale $\tau_6(T)$ but
also spin-waves on a more rapid time-scale, which we have not
described. The latter should contribute to the specific heat as in
normal two-dimensional antiferromagnets and give in particular a $T^2$
specific heat as observed experimentally.\cite{RamirezC} This is the
consequence of the two Goldstone modes associated with the selection
of a common plane (nematic broken symmetry).\cite{Ritchey}

We can make different assumptions regarding the time-scale of the
activated dynamics with respect to the observation time-scale.  If
$\tau_6(T) \gg \tau_{neut.}$, the system is trapped into a typical
3-coloring on the experimental time-scale. Still the averaged moment
is different from $S$ because of the rapid zero-point fluctuations of the spin-waves. One can
estimate that the effect of the two Goldstone modes is to reduce the
moment to $m=S-0.16$.\cite{cepasunp} For Cr$^{3+}$ ($S=3/2$), the
correction is small and cannot explain the small moment measured.

Suppose now that the hexagons still have a dynamics, as indeed
predicted for $T<T_g$. We found in this case that the frozen moment is
$\langle S_i \rangle^2 \approx 0.31$ (Fig.~\ref{Sloct}). Applying the
same zero-point motion reduction as above, we find
$0.31(1-0.16/S)^2=0.25$ which is close to the experimental frozen
moment.  The model therefore gives a fair account of the measured
frozen moment. The small static moment is not due to strong quantum
fluctuations but rather to the loop (hexagon) fluctuations.

\begin{figure}[h]
\centerline{
 \psfig{file=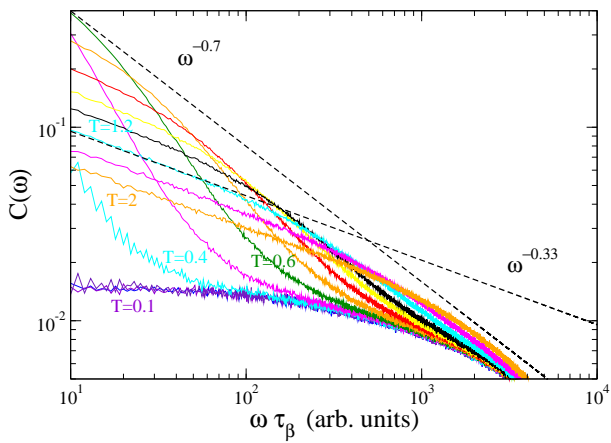,width=7cm,angle=-90}
}
\caption{Spectral function at different $T$ (dashed lines are $\omega^{-\alpha}$).}
\label{SlocFourier}
\end{figure}

To characterize the dynamics, we have computed the 
local dynamical response at different $T$
(Fig.~\ref{SlocFourier}). These are the Fourier
transforms of the autocorrelation functions given in Fig.~\ref{Sloct}.
At $T>T_g$, and low frequencies, we have
$C(\omega) \sim \omega^{-1/3}$ as the consequence of the universality
of the height model. However, this is valid over a limited range of
frequencies: in Fig.~\ref{SlocFourier}, the dashed lines give examples
of power-laws with exponents 0.33 and 0.7, for comparison (note that
all the curves are shifted horizontally by $1/\tau_{\beta}$). It is also
in fairly good agreement with the observed power law behavior in
neutron inelastic scattering on powders, $\omega^{-0.4}$ above the transition.\cite{Broholm,Lee,Mondelli} When $T$ is
lowered, the quasi-elastic peak corresponding to the frozen moment
develops. Note that the sum-rule $\int C(\omega) d\omega=1$ ensures that the
apparent loss of intensity at low temperatures in Fig.~\ref{SlocFourier} corresponds to a transfer into the
elastic peak.  Although the approach is different, we note that the exponent
is not far from that obtained by dynamical mean-field theory,
$\alpha \simeq 0.5$.\cite{Florens}

In summary, the model describes a dynamical freezing crossover into a
partially frozen phase and a small frozen moment in overall agreement
with the experiments. The broad neutron response is interpreted as the motion of loops above $T_g$. In the frozen phase, only the hexagons are
predicted to move (in addition to spin waves). They could possibly be
characterized by special magnetic form factors, as in
ZnCr$_2$O$_4$.\cite{Broholmnature}

\subsection{Volborthite Cu$_3$V$_2$O$_7$(OH)$_2$.2H$_2$O}

In volborthite,\cite{Hiroi} a freezing transition occurs at
$T_g \sim 1$K, with a finite static moment observed by NMR\cite{Bert,Yoshida} but no long-range correlations in neutron scattering.\cite{Nilsen} 
Volborthite is a slightly distorted kagome lattice
and there is some current debate as to whether the main magnetic
couplings are kagome like or more one-dimensional.\cite{Janson} We
will assume below that it can be viewed as a kagome antiferromagnet
and that the distortion is a small effect.
 
Below the transition, NMR revealed that the phase is heterogeneous
with a time-dependent lineshape, leading to distinguish between
``fast'' and ``slow'' (static) sites, either at small
fields\cite{Bert} or in a dinstinct phase,\cite{Yoshida} at larger
fields.\cite{Yoshida1}

These results resemble the dynamical heterogeneities found in the model below
$T_g$. We can make a more detailed comparison by computing the
distribution of fields. NMR was performed on vanadium nuclei which
seat at the centers of the hexagons.\cite{Bert,Yoshida} The nuclei see
effective fields averaged over the six sites $i_H$ of a hexagon $H$
(assuming for simplicity the same hyperfine coupling $A_{i_H}$),
\begin{equation}
\langle \mathbf{h}_H \rangle= \sum_{i_H=1}^6 A_{i_H} \frac{1}{t} \int_0^{t} dt'  \mathbf{S}_{i_H}(t'),
\end{equation}
which depend on the hexagon (inhomogeneous broadening). An average over the NMR time scale $t$ is
taken. In principle $t$ is much larger $\approx 10 -100 \mu$s than the
microscopic time-scales $\approx$ps, and $t$ can be taken to $+\infty$.
In systems with slow dynamics, NMR probes local trajectories averaged over $t$. The lineshape depends on $t$, thus providing information on the presence of dynamical heterogeneities. The
lineshape is related to the distribution function of field strengths
$P(h\equiv |\langle \mathbf{h}_H \rangle|)$, which we have calculated
in the present case.

\begin{figure}[h]
 \psfig{file=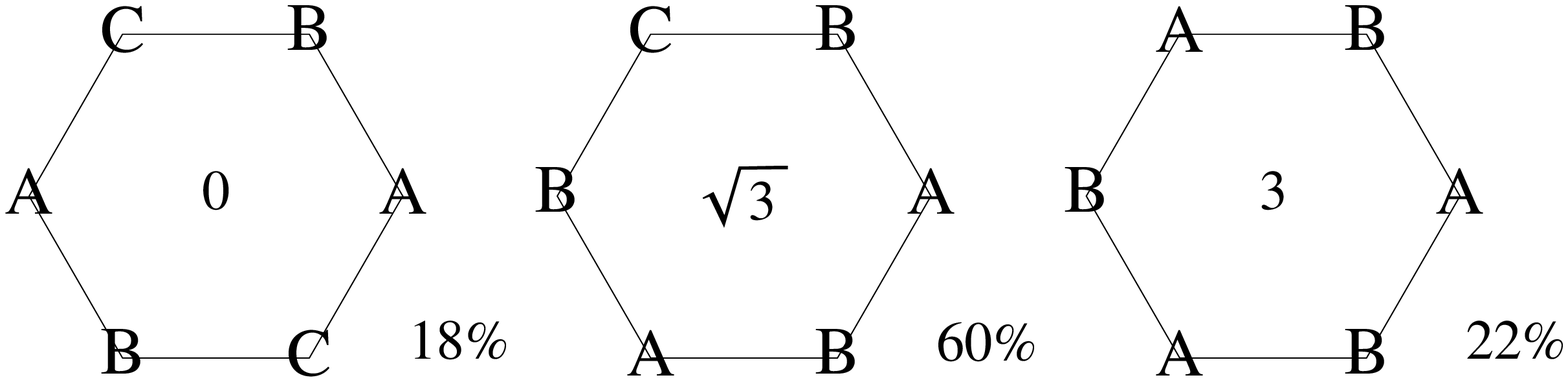,width=8cm,angle=-0} \\
\vspace{-.5cm}
 \psfig{file=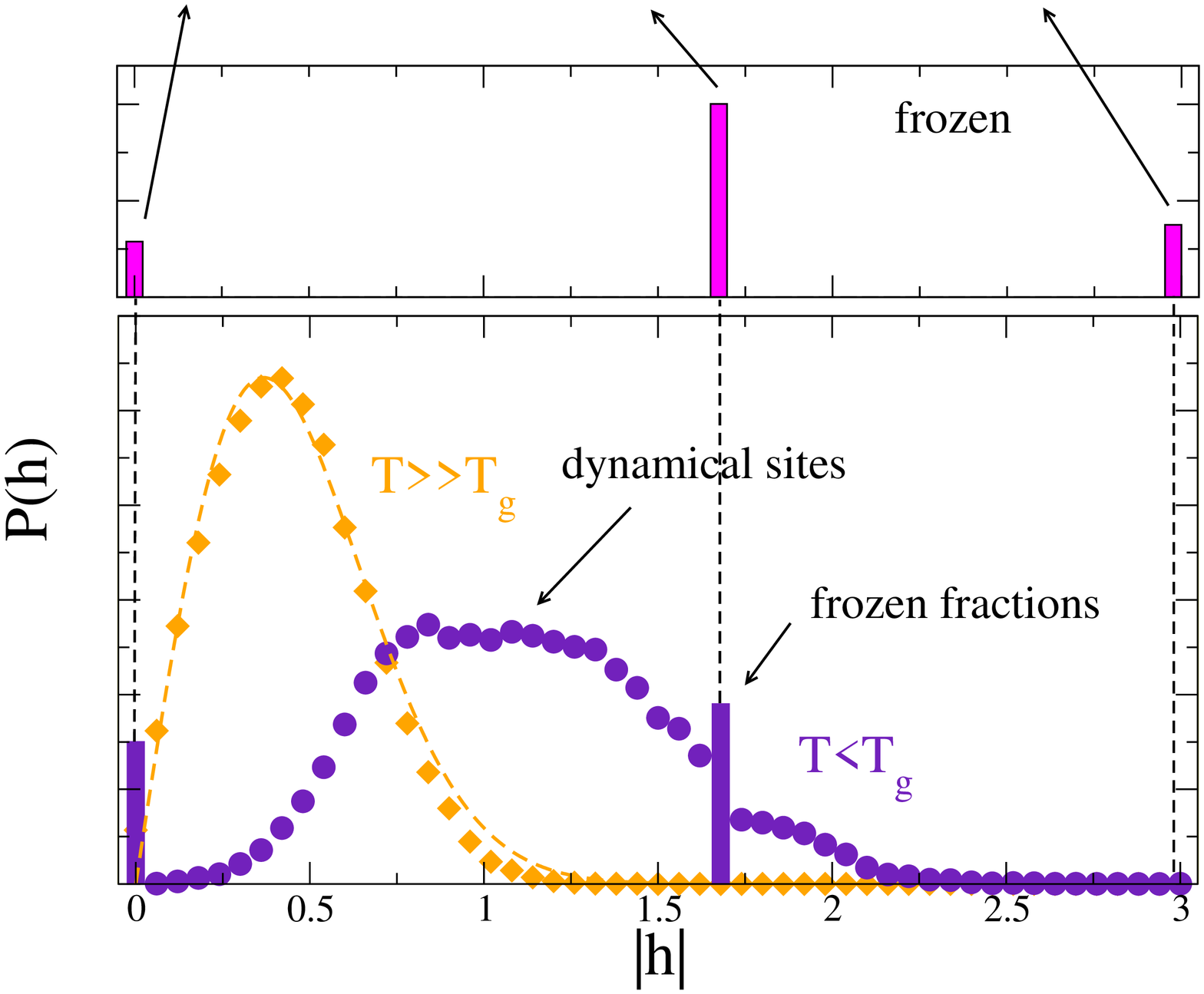,width=7cm,angle=-90}
\caption{Distribution of local field strengths at the centers of the hexagons. Evolution from high $T$ to $T<T_g$. A typical static state (no loop flip) is given for comparison (peaks at $0,\sqrt{3},3$, see top).}
\label{nmr}
\end{figure}

We expect different regimes, according to whether the NMR time-scale $t$ is shorter or longer than the characteristic time-scales of the dynamics, $\tau_{\beta}$ and $\tau_{\alpha}$. Note that since these describe activated processes, they may become much longer than the ps microscopic time at low temperatures.
\begin{itemize}

\item $t \gg \tau_{\alpha},\tau_{\beta}$. The system equilibrates on NMR time-scales, \textit{e.g.} at high-$T$. Every site has
  dynamics and summing random vectors (at 120$^o$ though) gives a
  gaussian distribution of fields (dashed line in Fig.~\ref{nmr}). For
  $t\rightarrow \infty$, summing local fields corresponds to a random walk and the typical strength $h \sim 1/\sqrt{t} \rightarrow 0$ since we have no external field.

\item $t \ll \tau_{\alpha}, \tau_{\beta} $. The system is completely frozen in a typical three-coloring state. Each nucleus sees a well defined static field. For a 3-coloring, there are only three possible field strengths at the center of the hexagon,  $h=0,\sqrt{3},3$ (see the configurations shown in Fig.~\ref{nmr}, top). Averaging over the uniform ensemble, we find three peaks with weight 18$\%$, 60$\%$, 22$\%$ (22$\%$ is the fraction of flippable hexagons). For comparison, the $Q=0$ antiferromagnetic state would have a single peak at $h=0$ with 100$\%$ of the hexagons and the $\sqrt{3} \times \sqrt{3}$ state a single peak at $h=3$.

\item $\tau_{\beta} \ll t \ll \tau_{\alpha}$. The system is
  out-of-equilibrium below $T_g$, by definition. The dynamical sites
  provide a time-dependent averaged field (broad part of the lineshape
  in Fig.~\ref{nmr}). The frozen sites inside the clusters provide a
  static field: we find two peaks at $h=0$ and $\sqrt{3}$ and no peak
  at $h=3$ which corresponds to the flippable hexagons. Although the
  local field does not change when they flip, the probability that
  they remain in a flippable configuration is small. Instead they move
  on the lattice and there are very few isolated flippable hexagons
  inside frozen clusters.  We further note that the static fields
  inside the frozen clusters show a ratio $P(0)/P(\sqrt{3}) \approx
  0.8$ much larger than that of a typical state $\approx 0.3$
  (Fig.~\ref{nmr}). This means that the frozen clusters resemble
  locally to the $Q=0$ state, the state with long linear winding
  loops, precisely those which do not flip.
\end{itemize}

Experimentally, in volborthite, the NMR lineshape consists of two
dynamically heterogeneous contributions at $T<T_g$.\cite{Yoshida1} A
slow ``rectangular'' shape was assigned to a static field (the
``rectangular'' shape arising from the powder convolution) and a fast
gaussian contribution to dynamical sites.\cite{Yoshida1} Similar results were obtained at low fields on lower-quality samples.\cite{Bert}

The low-field results can be compared with Fig.~\ref{nmr} (up to a
powder convolution). For $T<T_g$ we find two contributions: (i) a
static contribution coming from the frozen clusters and represented by
two peaks at $h=0$ and $h=\sqrt{3}$, the latter giving rise to a
rectangular shape in a powder sample. (ii) a dynamical part resulting
from the dynamical regions and giving rise to a broad
response. Here we do not have a single dynamical site but this is not
necessarily incompatible with the experiments because of the
difficulty of resolving different sites. We also note that the
respective contribution of both is smaller than the 50$\%$-50$\%$
observed experimentally,\cite{Yoshida1} but this depends on the
time-scale.

Assuming that the system had a $\sqrt{3} \times \sqrt{3}$ character
and that the static field was therefore due to the $h=3$ types of
hexagons, a small frozen moment of 0.41$\mu_B$ per site was
extracted.\cite{Bert} In the present model, a peak at $h=3$ is not
compatible with the existence of dynamical sites. Instead we assign
the experimental peak to the $h=\sqrt{3}$ frozen field. In this case,
instead of $m=0.41\mu_B$,\cite{Bert} the static moment is $m=0.41\mu_B
\times \sqrt{3} =0.71 \mu_B$ (as also proposed in
Ref.~\onlinecite{Wang} for different reasons), which is more
compatible with conventional on-site zero-point fluctuations,
$(1-0.16/S)\mu_B=0.68\mu_B$. Moreover, if we now calculate the total
frozen moment averaged over all sites as measured by neutrons (while
NMR sees the full local frozen moment), we would predict $m_{av}=0.56
\times 0.71\mu_B = 0.40\mu_B$.

We conclude that the present study gives a model for the phase
transition and the heterogeneous state observed in volborthite. Similarly to SCGO, it
gives an interpretation for the small moment observed for $T<T_g$: the
fluctuations of small loops reduce the averaged moment. The model suggests a
more precise picture of frozen clusters with an emergent length scale,
that can be further tested experimentally.

\subsection{Vesignieite Cu$_3$BaV$_2$O$_8$(OH)$_2$}

For vesignieite,\cite{Okamoto} $T_g=9$K, and the ground state is also
heterogeneous: approximately 50$\%$ of the sites (muon sites and nucleus sites) experience a static field.\cite{Quilliam,Colman} The
loss of 50$\%$ of the total intensity in NMR is due to the nuclei
which have a time-scale that cannot be detected, and therefore
reflects some dynamical heterogeneities in the local environment.
It would be inaccurate to consider that the 50$\%$ of the
observed intensity is due to the spins in the frozen clusters and the missing 50$\%$ due to the fast moving spins. It may well be that some dynamical sites of Fig.~\ref{nmr} are detected (this is in fact what we assumed for the
volborthite where 100$\%$ of the nuclei were detected). 
The fact that the fraction does not match the number of frozen sites of 12$\%$ is not therefore a serious drawback.  Alternatively
the fraction of frozen sites certainly depends on the
interactions. 
For Dzyaloshinskii-Moriya interactions which are present and may be rather strong,\cite{Quilliam} the frozen fraction will certainly increase because it favors the $Q=0$ state with long loops.

\subsection{Hydronium jarosite, (H$_3$O)Fe$_3$(SO$_4$)$_2$(OH)$_6$}

A freezing transition occurs at $T_g \sim$15K.\cite{jarosite} By
varying sample preparations, $T_g\sim$ 12K-18K and it appears to be
weakly sensitive to the Fe coverage in the range
92-100$\%$.\cite{Willspreparation} Neutron elastic scattering has
found short-range correlations of the $\sqrt{3} \times \sqrt{3}$ type,
 but no long-range order.\cite{WillsCT,Willsneutrons,Fak} The Heisenberg coupling is $JS^2=244$K
($S=5/2$),\cite{Fak} so that $T_g/JS^2=0.05$ which in terms of a
classical Heisenberg model means that the system should be in the
collective paramagnetic regime.\cite{Chalker} One clearly needs some additional
ingredients to explain the freezing transition.

Spin anisotropy is known to be present in a similar jarosite compound,
KFe$_3$(SO$_4$)$_2$(OH)$_6$, both single-ion easy-plane anisotropy
$DS^2\sim 30$K and a Dzyaloshinskii-Moriya interaction $|\tilde{D}|S^2
\sim 20$K explaining the excitation spectrum.\cite{Matan,Yildirim}
X-ray dichroism of the Fe$^{3+}$ ion also found a single-ion
anisotropy in good agreement with the above
figure.\cite{deVriesAnisotropy} In addition, in ordered jarosite
compounds, a second transition corresponding to the in-plane locking
of the spins occurs at $45-55$K.\cite{Frunzke} With these large values
in mind, we assume that the energy barriers of the model originate in
the anisotropy. In this case, we can predict $T_g$ and compare with
that obtained from ac-susceptibility measurements.\cite{Wills} Since,
in the model, we have $T_g \approx 0.3\kappa = 0.225DS^2$ for
$\tau_{exp}=10^3$s ($6 \times 10^{-3}$Hz), and $DS^2\sim 30-55$K, we
find $T_g \approx 7-12$K. Similarly, for $\tau_{exp}=80$ms (80Hz), we
find $T_g \approx 9-15$K. These estimates are a little smaller than
the experimental figures and depends more strongly on the measurement
frequency (the same distinction occurs in structural glasses between
``fragile'' and ``strong'' glasses).\cite{Wills} Moreover, by varying
synthesis conditions, $T_g$ was found to be correlated with the
distortion of the FeO$_6$ octahedra: the stronger the distortion the
larger the $T_g$.\cite{Bisson} Since the octahedron distortion implies
a linear change in the crystal field splitting, hence in the
single-ion anisotropy $D$, we expect indeed linear changes in $T_g
\approx D$, as observed experimentally.\cite{Bisson}

For $T<T_g$, an estimate of the frozen moment has been obtained by
$\mu$SR and amounts to 3.4$\mu_B$ compared with 5.92$\mu_B$ of the
Fe$^{3+}$ ion,\cite{HarrisonPhysicaB} so that $\langle S_i
\rangle=0.57$. It is not far from the present estimate
$0.56(1-0.16/S)=0.52$. However, it is surprising that similar values
were obtained in ordered jarosites.\cite{HarrisonPhysicaB}

For $T>T_g$, neutron inelastic scattering has been performed and
showed that the local response, $\chi''(\omega) \sim
\omega^{-0.68}$.\cite{Fak} At very low frequency, we have found
$\omega^{-1/3}$ but at larger frequencies it could be fitted by a
larger exponent (the second dashed line in Fig.~\ref{SlocFourier} corresponds
to $\alpha=0.7$). The agreement is therefore qualitative with a broad
increasing response by lowering the frequency (to be contrasted with
the flat response of a conventional two-dimensional antiferromagnet)
but a single exponent is not found.

To conclude, the present study suggests that $T_g$ in
(H$_3$O)Fe$_3$(SO$_4$)$_2$(OH)$_6$ is related to a dynamical freezing
into a heterogeneous state. The relevant energy scale here, contrary
to SCGO, is the anisotropy, as experimentally claimed.\cite{Bisson}
Below $T_g$, we expect a small frozen moment on average and a
\textit{persistent} dynamics of the hexagons, which distinguishes the
present transition from a complete dynamical arrest. More studies of the
low-temperature phase would be interesting.

\subsection{Other kagome compounds, competitions}

It is well known that not all kagome compounds have a freezing
transition and we briefly discuss some other compounds. Some have magnetic
long-range order, which is often accounted by additional spin
interactions. Others, such as the herbertsmithite compounds
ZnCu$_3$(OH)$_6$Cl$_2$\cite{Shores} and
MgCu$_3$(OH)$_6$Cl$_2$,\cite{Kermarrec} have no freezing transition
(unless an external field is applied\cite{Jeong}) and no long-range
order.\cite{MendelsJPhysSoc} The neutron inelastic response has no
clear energy scale in ZnCu$_3$(OH)$_6$Cl$_2$\cite{deVries} and is
fitted by a broad power law $\omega^{-0.67}$ at low enough
energy,\cite{Helton0,Helton1} with some similarity with that of SCGO
and the hydronium jarosite above $T_g$.  In the present model, one
would interpret this result as being in the phase above $T_g$, and the
neutron inelastic response agrees qualitatively with
Fig.~\ref{SlocFourier}. However, the reason why $T_g$ would be smaller
than the lowest temperatures reached experimentally, say 50 mK, is not
clear.  We have argued that $T_g$ is controlled by the anisotropy
(dynamically-generated or not), and the anisotropy is present in
ZnCu$_3$(OH)$_6$Cl$_2$.\cite{Zorko,cepasESR} Two important effects are
missing: it is known that antisite disorder is present,\cite{Olariu}
and that $S=1/2$ compounds have strong quantum effects with currently debated
quantum spin liquid phases if the anisotropy is sufficiently weak
(such a coupling may discriminate between different phases in $S=1/2$
compounds\cite{CepasDM}). It is therefore clear that competitions 
are important to account for all these phases.

\section{Conclusion}
\label{Conclusion}

We have described a simple spin model which has a dynamical glassy-like
freezing at a crossover temperature $T_g$, in absence of any quenched
disorder. 

The system evolves from a dynamically homogeneous phase with a single
time-scale ($T>T_d$) to a dynamically heterogeneous phase with two
time-scales ($T<T_d$). The first time-scale $\tau_{\beta} \sim
\tau_6(T)$ corresponds to the ``rapid'' degrees of freedom, the
shortest loops. The second time-scale $\tau_{\alpha}$ is associated
with the rearrangement of the ``frozen'' clusters. The frozen clusters
have a microscopic length-scale (they typically contain a few tens of
sites) but their rearrangement time is not controlled by the their size but by the size of the second shortest loops, $\tau_{\alpha}
\sim \tau_{10}(T)$. When $\tau_{\alpha}$
becomes longer than the experimental time-scale for $T<T_g$, the
system is out-of-equilibrium and glassy-like. The clusters contain spins that are frozen
on the experimental time-scale and realize a microscopic-scale
disorder.  In this case, the system has a finite (small) averaged frozen
moment but no true long-range order. We have explained that the
frozen moment is due, partly to the frozen clusters
themselves and, partly to dynamical regions where the spins are strongly
constrained by the frozen regions.

The phase space of the system appears to be organized in a partially
hierarchical manner with conserved quantities defining $\sim N$ basins
separated by infinite barriers (broken
ergodicity). Each basin was shown to further split into $e^{NS_c}$
sectors separated by finite barriers which trap the system in a
metastable state below $T_g$.  This macroscopic fragmentation of the
phase space corresponds to the local disorder induced by the
``frozen'' clusters. At $T_g$, the system has therefore some
``thermodynamic'' anomalies characterized by the loss of the
configurational entropy which we have calculated by finite-size
scaling, $S_c=0.082$ per site.

The system undergoes a glassy-like transition at $T_g$ because the
residual ``rapid'' degrees of freedom (the shortest
loops) only \textit{partially} reorganize the system. In a typical
state, the density of the shortest loops is not very small, but,
by effectively attracting each other, they form aggregates and voids
(micro phase separation), the latter regions being, hence,
frozen. Some details, as to what their density is, or how they precisely interact, certainly depend on the
system and the model, but the mechanism we have presented here is
rather clear: the strong local correlations generate slow extended
degrees of freedom which, since they are correlated and attract each other, ``phase-separate''
in dense active regions and void inactive regions.

 Several aspects of the degenerate model are simply
 \textit{assumed}. We have assumed the absence of long-range order by
 considering degenerate states (Eq.~\ref{Ei=0}), and an activated
 relaxation time (Eq.~\ref{tau}); hence, not surprisingly, the
 dynamics is slow. We have discussed in section \ref{micro} why both
 assumptions may be approximately realized in microscopic models with
 continuous degrees of freedom. We argued that the origin of the
 energy barriers is the partial order-by-disorder, i.e. the barriers
 are dynamically generated by the rapid spin-waves, or by an explicit
 anisotropy arising from the spin-orbit coupling. The degeneracy
 (Eq.~\ref{Ei=0}) is in general not exact and lifting it favors a
 ``crystal'' state in the energy landscape without modifying -if it
 remains sufficiently small- the dynamical aspects we have described.

We have compared the results with the experiments on the kagome
compounds. The present study gives a model for the spin freezing
observed at $T_g$ and provides an interpretation for the nature of the
low-temperature phase. The picture of the ``frozen'' phase that
emerges is that of an heterogeneous state with dynamical and frozen
regions. The weak measured frozen moment is interpreted as the
consequence of the remaining dynamics of the shortest loops and its
strength is close to what is measured in the experiments.  While in
magnets in general, the on-site moment is reduced by the small
oscillations around the ordered state (spin-waves), here the main
effect is argued to be the large-amplitude motion of the shortest
loops. The short loop fluctuations do not fully destroy the moment for
$T<T_g$ but their presence is in agreement with the persistent
fluctuations observed by different experimental techniques (neutrons,
$\mu$SR, NMR). In particular, the observation in NMR of nuclei with different time-scales is consistent with the heterogeneous picture of the dynamics proposed here.  In conventional magnets, the thermal excitations of the
spin-waves destroy the on-site magnetization. Here, one needs longer
loops, that are thermally excited only for $T>T_g$. These fluctuations
give a spectral response that obeys a power-law $\omega^{-1/3}$ in the
small energy limit, very different from that of conventional magnets
(flat response in two dimensions). A broad power-law response is
indeed observed experimentally in neutron inelastic
scattering. Although the exponent seems to be underestimated, the
experiments may not have had access to the low-energy limit or the
exponent may be inacurately predicted because of the interaction between the spin-waves and
the discrete modes.
In the paramagnetic phase, the model has algebraic spatial
correlations at equilibrium ($T>T_g$), a feature that is not observed
in elastic neutron scattering.  We believe that this is not
redhibitory, for the spin freezing we have described is not related to
the long-distance behavior. In two spatial dimensions, the correlation
length is always finite at finite temperatures.\cite{Chakra}
Furthermore, the chemical disorder is present to an amount which is
difficult to quantify and which has been completely neglected here.

The energy scale that governs the freezing temperature $T_g$
is argued to be $J$ in the small anisotropy limit (dynamically generated barriers), $T_g =0.04 JS$ and it crosses
over to $T_g= 0.225DS^2$ in the strong anisotropy limit, typically if
$D/J>0.18/S$.  This led us to a tentative classification, where SCGO
is in the small anisotropy limit and
(H$_3$O)Fe$_3$(SO$_4$)$_2$(OH)$_6$ in the strong anisotropy
limit. This is clearly a different interpretation from that of
chemical disorder, where $T_g$ is governed by the amount of
disorder.\cite{Saunders}

In order to disentangle intrinsic effects from the effects of chemical
disorder, one can test the present theory, in particular by
characterizing experimentally the active magnetic degrees of freedom, for
instance by neutron form factors,\cite{Broholmnature} or by inferring
the nanoscopic size of the frozen clusters.

\acknowledgments

O.~C. would like to thank J.-C. Angl\`es d'Auriac, F. Bert, L. Cugliandolo, B. Dou\c cot, B. F\aa k,  D.~Levis, C. Lhuillier, P. Mendels, H. Mutka, G. Oshanin, and J. Villain for discussions and especially A. Ralko for continuing collaboration. B.~C. would like to thank M. Taillefumier, J. Robert, C. Henley and R. Moessner for discussions and collaboration on related projects. O.~C. was partly supported by the ANR-09-JCJC-0093-01 grant.


\begin{thebibliography}{99}
\bibitem{cautiousname} Note that ``dynamical heterogeneity'' is used
  in a sense slightly different from that used in structural
  glasses,\cite{reviewglass} since dynamical correlations are not
  probed in NMR. What it literally means is that there are sites with
  different dynamics.
\bibitem{noteanisotropy}
Such axis result
from crystal fields and must be compatible with the
local symmetry (\textit{e.g.} the axis
pointing towards the center of the hexagon is in general different
from the other two, $E_1 \neq E_2 =E_3$).
\bibitem{notedeconfinement} The equivalent quantum model at
$T=0$ can be written down in 2+1 dimensions; the locking potential (Coulomb charges) is
relevant in the renormalisation group sense and the system was
conjectured to order in the $\sqrt{3} \times \sqrt{3}$ ground
state,\cite{Read} which was recently confirmed numerically by using a
microscopic model.\cite{cepasralko} Dimensional reduction leads to
Eq.~\ref{F} above a deconfinement crossover
temperature.\cite{cepasralko}
\bibitem{lr} In terms of spin structures indeed (see
  Ref.~\onlinecite{Harris} for the definitions of the structures) the
  $\sqrt{3} \times \sqrt{3}$ state is the one which maximizes the
  number of flippable hexagons (staggered chiralities), while the
  $Q=0$ state (all triangles in the same ABC state with uniform
  chiralities) has infinite loops scaling with the linear size of the
  system. The first (resp. second) is favored by a ferromagnetic
  (resp. antiferromagnetic) second-neighbor interaction $J_2$ (here $J_2=0$).
\end{thebibliography}
\end{document}